\documentclass[conference]{IEEEtran}
\IEEEoverridecommandlockouts

\usepackage{graphicx}
\usepackage[hyphens]{url}
\usepackage{graphicx} 
\usepackage{amsmath}
\usepackage{listings}
\usepackage{color}
\usepackage{listings}
\usepackage{caption}
\usepackage{subcaption}
\usepackage{xltabular}
\usepackage{longtable}
\usepackage{multirow}
\usepackage[numbers]{natbib}
\usepackage[bookmarks=false]{hyperref}
\usepackage{textcomp}
\usepackage{nameref}
\usepackage{algorithm}
\usepackage{textcomp}
\usepackage{algpseudocode}

\usepackage{tikz}
\usetikzlibrary{positioning}
\definecolor{dkgreen}{rgb}{0,0.6,0}
\definecolor{gray}{rgb}{0.5,0.5,0.5}
\definecolor{mauve}{rgb}{0.58,0,0.82}

\def\BibTeX{{\rm B\kern-.05em{\sc i\kern-.025em b}\kern-.08em
    T\kern-.1667em\lower.7ex\hbox{E}\kern-.125emX}}
    
\usepackage{tikz} 
\usepackage{textcomp} 
\usepackage{hyperref} 
\usepackage{lipsum} 
\newcommand\copyrighttext{  
\footnotesize \textcopyright 2021 IEEE. Personal use of this material is permitted. Permission from IEEE must be obtained for all other uses, in any current or future media, including reprinting/republishing this material for advertising or promotional purposes, creating new collective works, for resale or redistribution to servers or lists, or reuse of any copyrighted component of this work in other works.   DOI: \href{https://doi.org/10.1109/QRS54544.2021.00054}{10.1109/QRS54544.2021.00054}
} 
\newcommand\copyrightnotice{
\begin{tikzpicture}[remember picture,overlay] \node[anchor=south,yshift=10pt] at (current page.south) {\fbox{\parbox{\dimexpr\textwidth-\fboxsep-\fboxrule\relax}{\copyrighttext}}};
\end{tikzpicture}%
}    

\begin{document}

\title{\textit{Automated Cause Analysis of Latency Outliers Using System-Level Dependency Graphs}}

\author{\IEEEauthorblockN{Sneh Patel}
	\IEEEauthorblockA{\textit{sp18oo@brocku.ca} \\
		\textit{Brock University}\\
		St. Catharines, ON Canada \\
		L2S 3A1}
	\and
	\IEEEauthorblockN{Brendan Park}
	\IEEEauthorblockA{\textit{bp18ul@brocku.ca} \\
		\textit{Brock University}\\
		St. Catharines, ON Canada \\
		L2S 3A1}
	\and
	\IEEEauthorblockN{Naser Ezzati-Jivan}
	\IEEEauthorblockA{\textit{nezzati@brocku.ca} \\
		\textit{Brock University}\\
		St. Catharines, ON Canada \\
		L2S 3A1}
	\and
	\IEEEauthorblockN{Quentin Fournier}
	\IEEEauthorblockA{\textit{quentin.fournier@polymtl.ca} \\
		\textit{Polytechnique Montréal}\\
		Montreal, QC Canada \\
		H3T 1J4}
}

\maketitle

\begin{abstract}
    Detecting performance issues and identifying their root causes in the runtime is a challenging task. Typically, developers use methods such as logging and tracing to identify bottlenecks. These solutions are, however, not ideal as they are time-consuming and require manual effort. In this paper, we propose a method to automate the task of detecting latency outliers using system-level traces and then comparing them to identify the root cause(s). Our method makes use of dependency graphs to show internal interactions between threads and system resources. With these graphs, one can pinpoint where performance issues occur. However, a single trace can be composed of a large number of requests, each generating one graph. To automate the task of identifying outliers within the dataset, we use machine learning density-based models and statistical calculations such as $Z$-score. Our evaluation shows an accuracy greater than 97\% on outlier detection, making them appropriate for in-production servers and industry-level use cases.
\end{abstract}

\copyrightnotice

\section{Introduction}

Performance issues such as unexpected latency can negatively affect a program. During the development phase, debuggers and profilers help detect and resolve potential performance issues originating from bugs or poor design choices. However, after the application has been released, new variables that the developer may not have expected, such as different workloads or a new piece of hardware, may cause performance issues. Investigating latency issues of a released application and identifying their potential root causes is known to be a difficult task. One method often used for performance issues is execution trace-based runtime analysis.

Analyses are conducted either on-CPU or off-CPU depending on the type of issue~\citep{ezzati2020depgraph, zhou2018wperf}. When the analysis is conducted on-CPU, any threads running on the CPU can be analyzed. As a consequence, on-CPU analysis is only capable of detecting issues that occur during the execution time on the CPU. Such analysis is limited as it cannot detect performance issues related to thread scheduling and thread interactions. These performance anomalies can be detected with an off-CPU analysis. Notably, off-CPU analysis can detect issues related to hardware, I/O, network timers, and thread-waiting. For instance, let us consider a Java program that performs numerical computations with integers read from a local text file. In this situation, the on-CPU analysis can detect performance issues related to the calculation, while the off-CPU analysis can detect performance issues associated with reading the file. The proposed method in this paper is about analyzing latency outliers using off-CPU analysis. We collect runtime execution data by tracing the operating system kernel. We then convert the collected trace data into dependency graphs to model the internal structure of an execution. We base our latency and root cause analyses on these graphs.

A trace represents all the functions and operations accessed at the layer traced, either the kernel or the user space, during a program's execution. Each time that a tracepoint\footnote{A tracepoint is a tracing macro added to the source code or binary code} is encountered during runtime, an event is generated and added to the trace. An event is composed of a name, a timestamp, and possibly many other arguments. For example, the following event (\texttt{thread1}, \texttt{syscall\_open}, \texttt{file1}, \texttt{cpu0}, \texttt{t1}) corresponds to a file open system call, running on core 0, called on thread 1 at time \texttt{t1}. The Linux kernel already contains read-to-use tracepoints, making kernel-level tracing straightforward. Furthermore, kernel trace data includes all the interactions of a program with other threads/processes running simultaneously on the system and the system resources accessed during the execution of a program. This makes kernel tracing an excellent data source to investigate execution issues such as performance and latency issues. However, given the size of the collected trace data, they can be hard to analyze and may introduce a high overhead to the system. Therefore, there is a real need to introduce an efficient and automatic analysis tool for massive traces such as those collected at the kernel level.

In this paper, we propose a method to meet that demand. Our method automates the task of finding latency outliers in system-level traces. Additionally, the method is able to identify the root cause of latency issues by comparing individual outliers with clusters of normal executions to find shared and distinct behaviours between them. Automating the task of identifying outliers reduces the manual effort required to analyze a trace dataset. To automate the task, we first filter out the parts of the trace that do not correspond to the program. Using the filtered trace data, we then construct Waiting-Dependency Graphs (DepGraphs)~\citep{ezzati2020depgraph}, displaying the internal interactions between threads and system resources. Next, we use graph embedding techniques to convert the DepGraphs into fixed-size vectors compatible with most off-the-shelf machine learning algorithms. We use the vectors in density-based clustering models to detect outliers. Finally, we use our proposed merging and comparison algorithms to find the areas contributing to the latency issues. Our work in this paper expands upon what was presented by \citet{ezzati2020depgraph}. The authors discuss converting system-level traces into DepGraphs to reduce the effort required to analyze raw trace data. In this paper, our main contribution is the automation of analyzing DepGraphs for the outlier detection as well as clustering them to use in discovering the root cause of the outliers.

The remainder of this paper is organized as follows. Section~\ref{related_work} reviews the related work. Section~\ref{methodology} explains our proposed method for outlier detection and root cause analysis. Section~\ref{general_disscussion} provides an evaluation regarding the automated analysis along with a use case. Section~\ref{conclusion} summarizes the results and concludes this paper.

\section{Related Work}
\label{related_work}

Performance is an essential consideration for many developers when designing their programs. For this reason, developers often spend a significant fraction of their time investigating latency issues and optimizing their code. This section looks at some of the existing methods designed to detect and analyze performance issues.
 
The first work is GAPP~\citep{nair2020gapp} which uses weighted criticality measures to identify different serialization bottlenecks in parallel Linux applications caused by an imbalance or shared resource contention. GAPP can reveal a wide range of bottleneck-related performance issues, and the authors show how the extended Berkeley Packet Filter (eBPF) tracer can make a lightweight-based profiler. When used in combination with the bottleneck detector, it will pinpoint the lines of code responsible for the bottleneck. However, profilers are known to be ineffective because they operate by averaging the metrics, which may hide outliers~\citep{fournier2019automatic}.

Another method for bottleneck detection is using a causal profiling technique. These techniques discover bottlenecks and display the effects on the program were it to be optimized. COZ~\citep{curtsinger2015coz} is a causal profiler that uses a virtual speed up to perform causal profiling without actually optimizing the program's code. However, limitations to causal profiling do exist. For instance, causal profiling only applies to user space as it affects only the thread level. To address this issue, \citet{ahn2020scoz} introduced SCOZ, a system-wide causal profiler. Nevertheless, SCOZ is also a profiler and thus ineffective when it comes to outliers.

Similar to our proposed method, \citet{inagaki2019profile} uses thread graphs to detect layered bottlenecks. They presented a method to build a model that detects layered bottlenecks by profiling threads and their dependencies in the target system. This method is also capable of analyzing third-party components. They detect layered bottlenecks by hierarchically traversing the path with the most significant threads in their thread dependency graph. One limitation that remains in this work is the manual effort required from the user, which our method seeks to reduce.

\citet{10.1145/956750.956831} introduces two new techniques for graph-based anomaly detection. They present an algorithm called Subdue that can find repetitive patterns (substructures) within graphs. For their first approach, the authors use Subdue to find patterns within the dataset, and then any substructure that is opposite of the patterns is declared to be an anomaly. For their second approach, the authors separate the graph into distinct structures and compare them to each other to find anomalous subgraphs. The limitation of these methods is that Subdue can be very biased towards smaller graphs.

\citet{lane1997application} propose machine learning algorithms to detect anomalous behaviour within their program. Their system learns a user profile and subsequently applies it to detect anomalous behaviours. The program compares the current behaviours with the user profiles using a similarity measure. To determine whether the current behaviour is abnormal, input tokens are divided into fixed-size segments and compared using a similarity measure. Using this technique, the authors aim to detect intrusion attacks done by an outsider automatically.

Critical paths~\citep{doray2016diagnosing} can also be utilized to resolve performance problems. A critical path can show the state of threads' executions at different points of time to the user. A drawback of critical paths is that they only show a limited view of the execution rather than the complete process. For example, the critical path of a thread that is waiting for another thread to finish would not display the details of that other thread. Dependency graphs, proposed by \citet{ezzati2020depgraph} and adapted in our work, can show all thread interactions and help identify bottlenecks.

A common factor with existing tools for performance anomaly detection is that they require some degree of manual labour, making the analysis time-consuming and tedious. Our proposed method uses Waiting-Dependency Graphs (DepGraphs) to illustrate the interactions between different threads and system resources (i.e., disk, CPU, network, etc.). Unlike \citet{ezzati2020depgraph}, however, we aim to avoid manual analysis of these DepGraphs. Instead, we want to automate the process of detecting outliers within a dataset of DepGraphs. To do so, we utilize machine learning algorithms over the DepGraph data to identify outliers. After detecting outliers, we can compare them with a group of normal DepGraphs to identify the root causes of the anomalies.

\section{Methodology}
\label{methodology}

This section provides an in-depth explanation of the proposed anomaly detection and root cause identification method. Using a kernel-level trace dataset, we extracted multiple requests to convert them into DepGraphs. Each DepGraph shows an overall view of the request's internal behaviours, which can then pinpoint latency issues. However, the analysis becomes time-consuming as the number of DepGraphs increases. Our proposed approach converts the DepGraphs into graph embeddings that are later used in machine learning clustering and outlier detection algorithms to detect and investigate outliers automatically. After identifying the outliers, we compare them with clusters of normal executions to pinpoint the exact reasons behind the issues. In the following paragraphs, we explain each step in more detail.

\subsection{System-Level Data Collection}
\label{data}

We use a low-overhead open-source Linux tracing tool called the Linux Tracing Toolkit: next generation (LTTng~\citep{desnoyers}) to collect kernel-level trace data from a Linux operating system. The collected trace includes the full execution details of all active threads of the system. A thread may handle several tasks in its entire lifetime. We are, however, interested only in some aspects of a group of active threads. We first extract the requests (e.g., web requests) handled by a web server thread (or a group of similar threads, say Apache threads) from its kernel execution trace data, and then we construct a DepGraph for each request. This paper relies on the algorithm presented by \citet{ezzati2020depgraph} to extract the requests and construct the DepGraphs from trace data. Although this paper focuses on web requests, our method is generic enough to apply to any execution (e.g., a button click, a compiler event, a graphical-view rendering, or a network service).

\subsection{Waiting-Dependency Graphs}
\label{depgraph}

A Waiting-Dependency Graph (DepGraph)~\citep{ezzati2020depgraph} is a directed graph whose nodes are labelled with the name of the thread, system call, or other execution states and with their runtime duration in milliseconds. In this graph, edges indicate the percentage of time the source node spends waiting for the destination node. DepGraphs show a visual representation of a request's internal execution breakdown. Including the aggregated duration of a thread running on the CPU and the aggregated duration of a thread waiting for a system resource (e.g., disk, network, timer, CPU).

In previous works, developers would use critical paths to observe how long each thread's state takes~\citep{fournier2019automatic}. A critical path is a tool to display the execution states of a given thread at any particular time. However, a critical path cannot generally show why a thread is in a waiting state or how long a thread has been waiting for a resource. One would need to do a manual investigation to find why a thread is in a waiting state. If only a few threads exist investigating them is an easy task to do. However, in a real-world scenario, there are usually a large number of threads, making them tedious to investigate. In contrast, DepGraphs display the overall breakdown of waiting dependencies that cause latency issues. In essence, DepGraphs are the aggregation of all interacting threads' critical paths that contribute to handling a request. Consider, for example, a web server thread that is communicating with a database thread to handle a web request. In this case, the DepGraph of the web request is constructed by merging the critical paths of the two threads. For more information regarding the construction of DepGraphs we refer the reader to the original paper by~\citet{ezzati2020depgraph}.

\begin{figure*}[!htb]
	\centering
	\begin{subfigure}[b]{0.49\linewidth}
		\centering
		\includegraphics[width=0.75\linewidth]{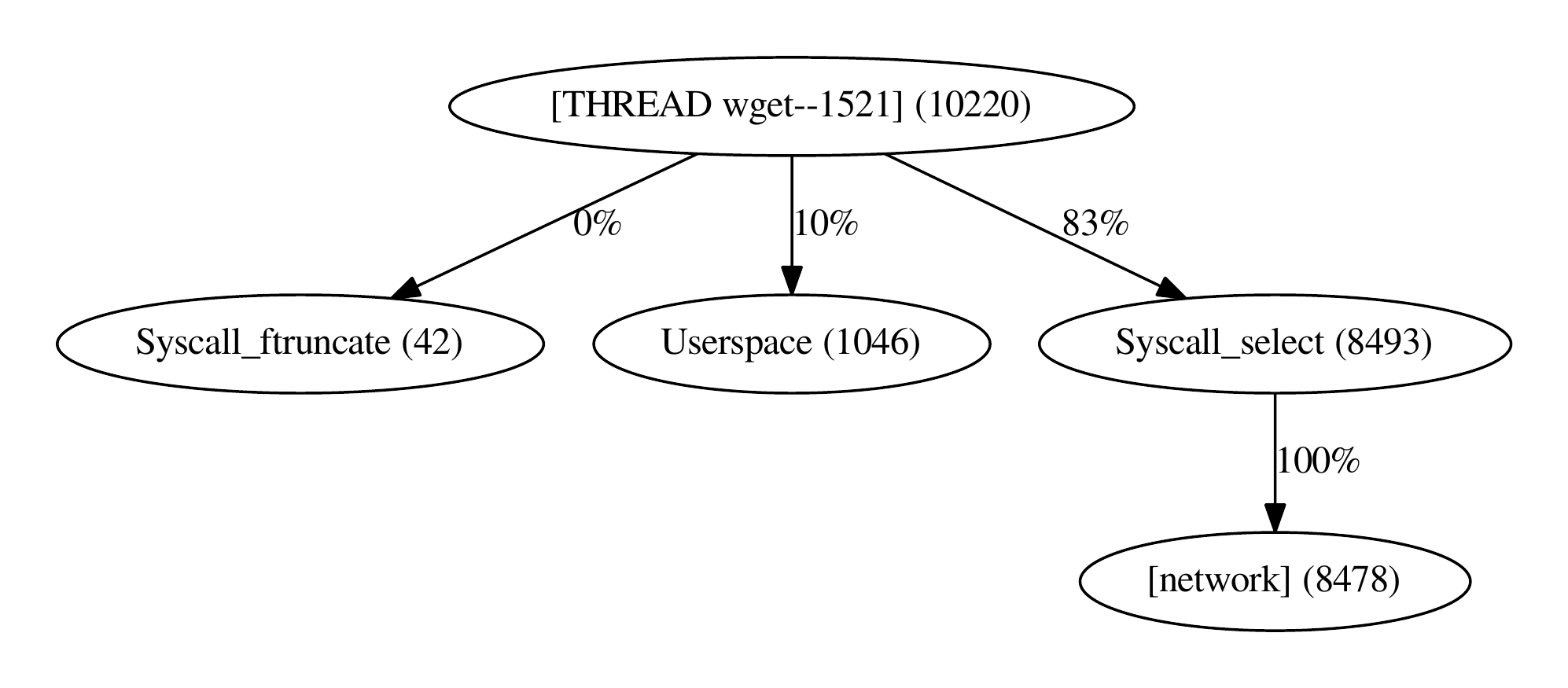}
		\caption{Client DepGraph of the first request.}
		\label{wget1}
	\end{subfigure}\hfill
	\begin{subfigure}[b]{0.49\linewidth}
		\centering
		\includegraphics[width=0.75\linewidth]{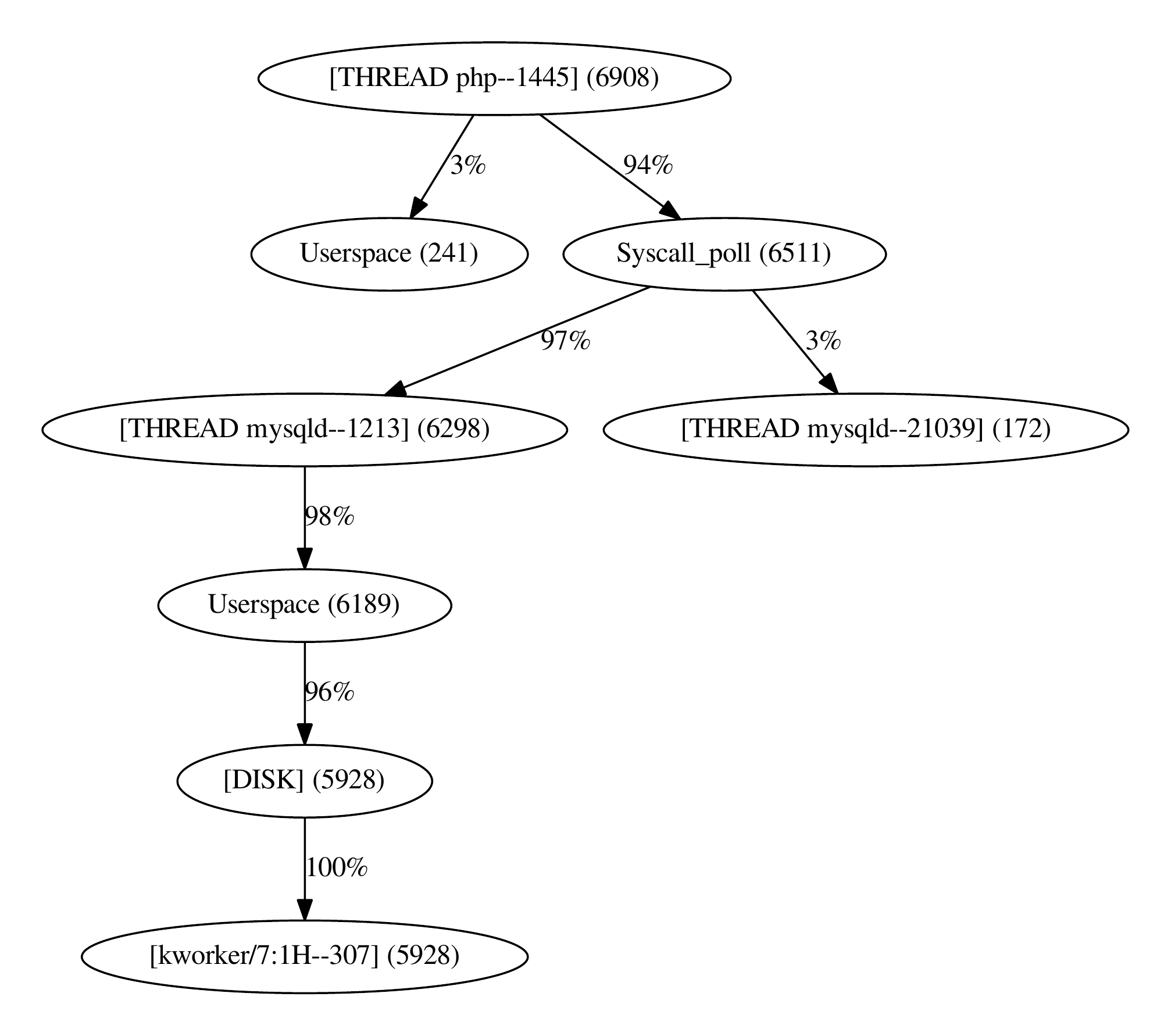}
		\caption{Server DepGraph of the first request.}
		\label{php1}
	\end{subfigure}
	\begin{subfigure}[b]{0.49\linewidth}
		\centering
		\includegraphics[width=0.75\linewidth]{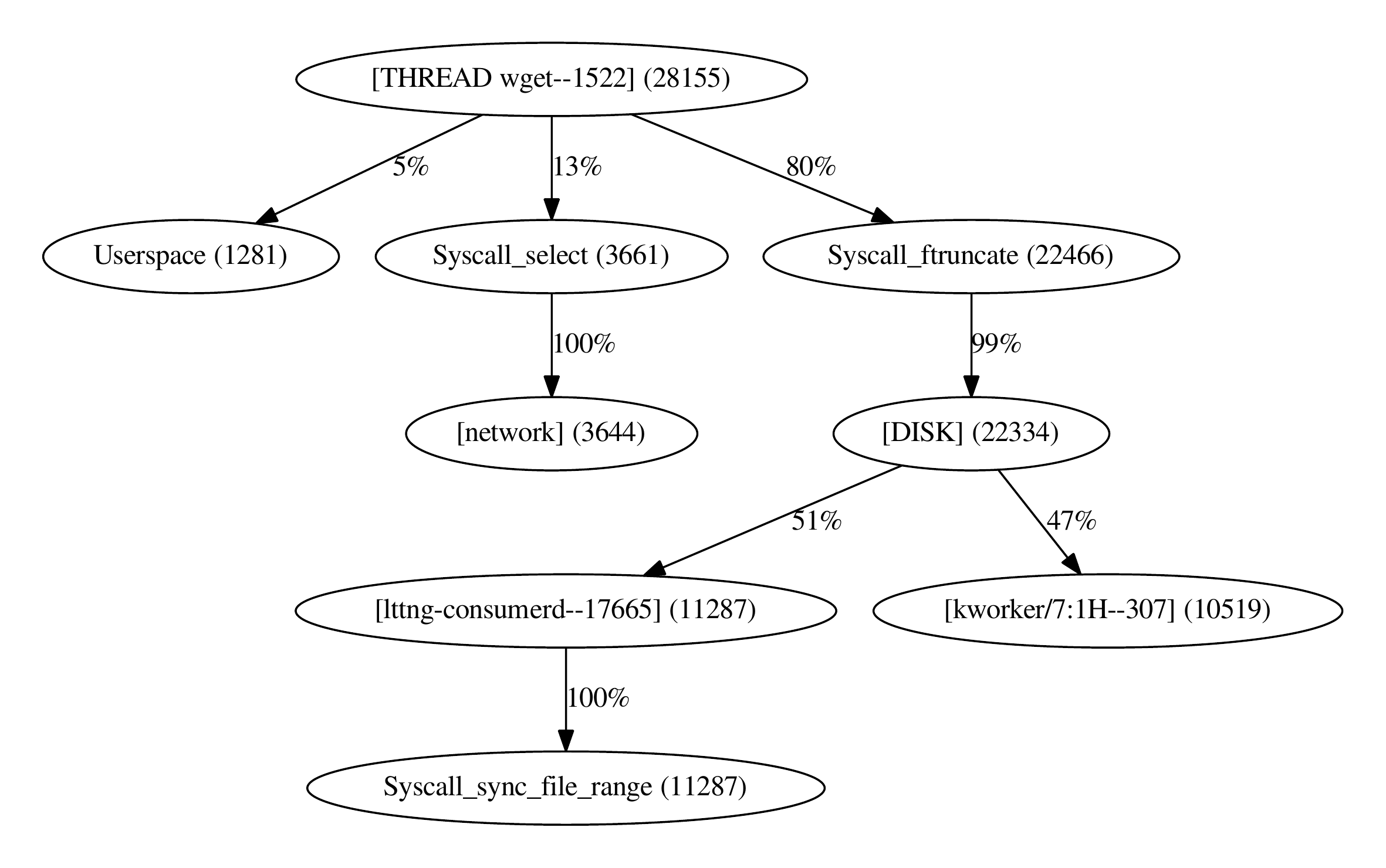}
		\caption{Client DepGraph of the second request.}
		\label{wget2}
	\end{subfigure}\hfill
	\begin{subfigure}[b]{0.49\linewidth}
		\centering
		\includegraphics[width=0.75\linewidth]{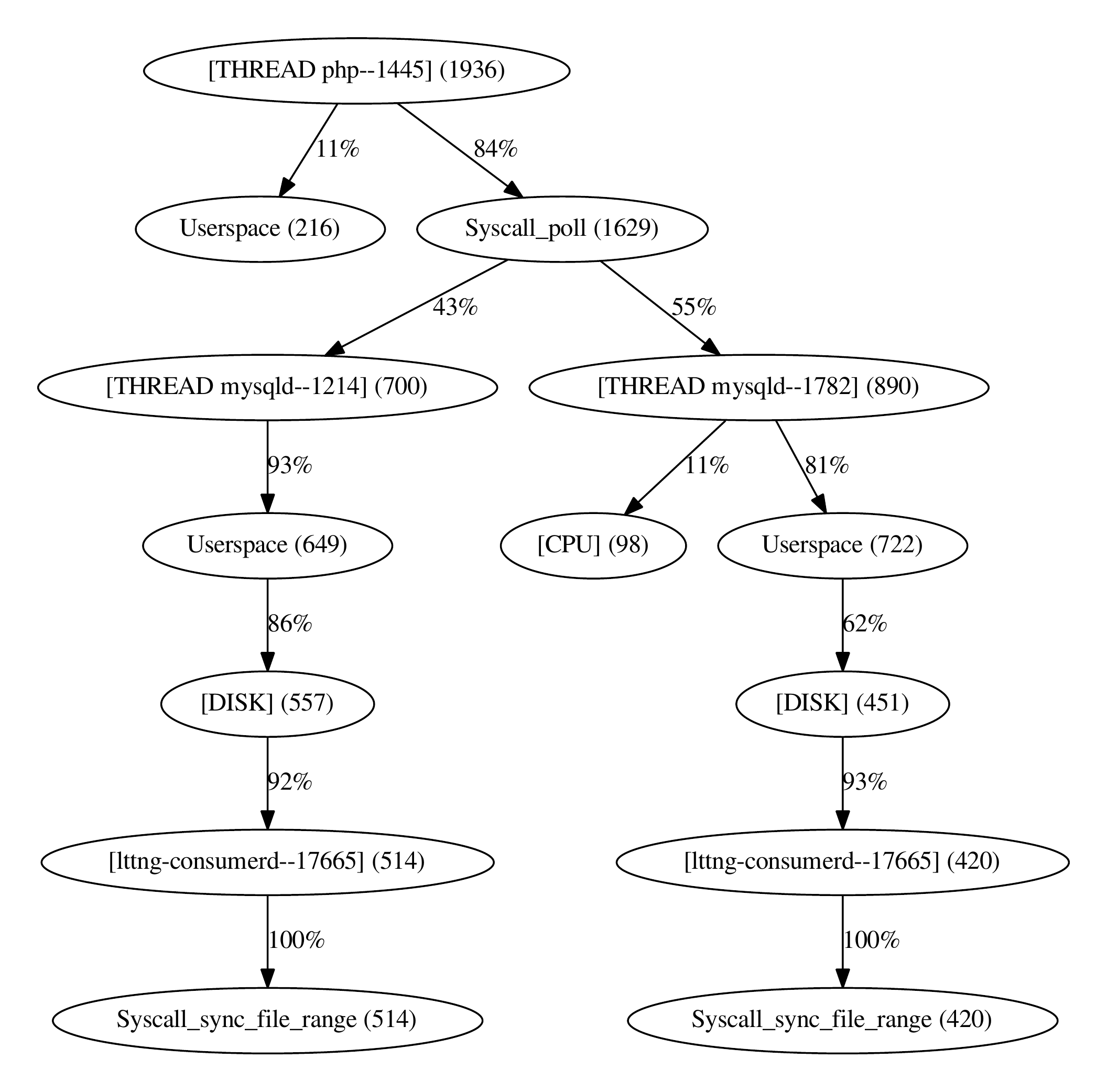}
		\caption{Server DepGraph of the second request.}
		\label{php2}
	\end{subfigure}
	\caption{The Waiting-Dependency Graphs of two \texttt{wget} requests handled by PHP web server. (Top) The first request behave as expected. (Bottom) The second request is slower than the first one, although MySQL cached the query.}
	\label{wgetexperiments}
\end{figure*}

DepGraphs allow finding latency issues and their root causes by comparing the graphs of the different normal and abnormal requests executions. To illustrate how DepGraphs can be used to find anomalies' root cause, let us consider the example of a web server that runs a PHP script to print all the table records from a MySQL database. The script repeats this database access three times. The command-line web client \texttt{wget} was used to send two consecutive requests to the server. In this example, one would expect the second call to be faster due to client and server caching mechanisms. However, we observed the opposite to be true. For this scenario, we have made four DepGraphs (Figure~\ref{wgetexperiments}): one for each client and server execution. Note that all edges with less than 3\% have been removed for readability since the edges with low values provide no significant latency to the execution. In Figure~\ref{wget1}, the DepGraph of the client execution of the first request shows that 83\% of the \texttt{wget} thread's execution was spent waiting for the network. In Figure~\ref{php1}, the DepGraph of the first request's server execution, we observe what occurred on the server during that time. Note that although \texttt{syscall\_ftrucntate} takes less than 3\% of the time, we preserved it for comparison with the second request. Figure~\ref{php1} shows that the majority of the server time was spent waiting for the MySQL database to handle the queries. For the second request, from the server's DepGraph shown in Figure~\ref{php2}, we can see that the database requests only take 700ms (compared to the 6298ms of the first request). Such a decrease in response time indicates that database files/queries are (probably) cached in memory from the first call, making the second database call much more efficient. Therefore, the web/database server is not the bottleneck. The DepGraph of the second request's client execution (Figure~\ref{wget2}) explains why the second request is slower than the first request. The latency is mainly due to a new state in the execution of the \texttt{wget} client called "Syscall\_ftruncate" where the \texttt{wget} waits for the disk to write something while already busy serving tasks from two other threads (\texttt{[kworker/7:1H--307]} and \texttt{[lttng-consumerd--17665]}). This explains why the total time of the second request has increased by almost three times compared to the first request.

As illustrated by the example, the analysis of DepGraphs, although very useful, is manual and requires significant human labour to identify issues. In order to reduce the manual labour behind analyzing these DepGraphs, we convert them into fixed-size vectors and analyze them with machine learning algorithms.

\subsection{Graph2Vec}
\label{graph2vec}

The proposed method is to analyze DepGraphs with machine learning algorithms automatically. Most off-the-shelf machine learning algorithms require real-value vectors as input; thus, one must embed the graphs to use them as input. Graph embedding is the approach of transforming nodes, edges, and other graph features into a vector space. For our proposed method, we used a technique introduced by \citet{DBLP:journals/corr/NarayananCVCLJ17} called Graph2Vec. Given a set of unidirectional graphs \(\textbf{G} = \{G_1...G_n\} \), Graph2Vec is able to learn graph embeddings of any given size that contains information pertaining to the graph's topology. Given a dataset of graphs, Graph2Vec attempts to explore all the nodes of their sub-graph up to a certain degree. The technique follows Doc2Vec \citep{DBLP:journals/corr/LeM14} Skip-gram training process. It first maps random numerical values to the vector spaces, and then using the Weisfeiler-Lehman (WL) relabeling strategy, they are trained and refined over several epochs.

\begin{figure}[htb]
	\centering
	\includegraphics[width=0.7\linewidth]{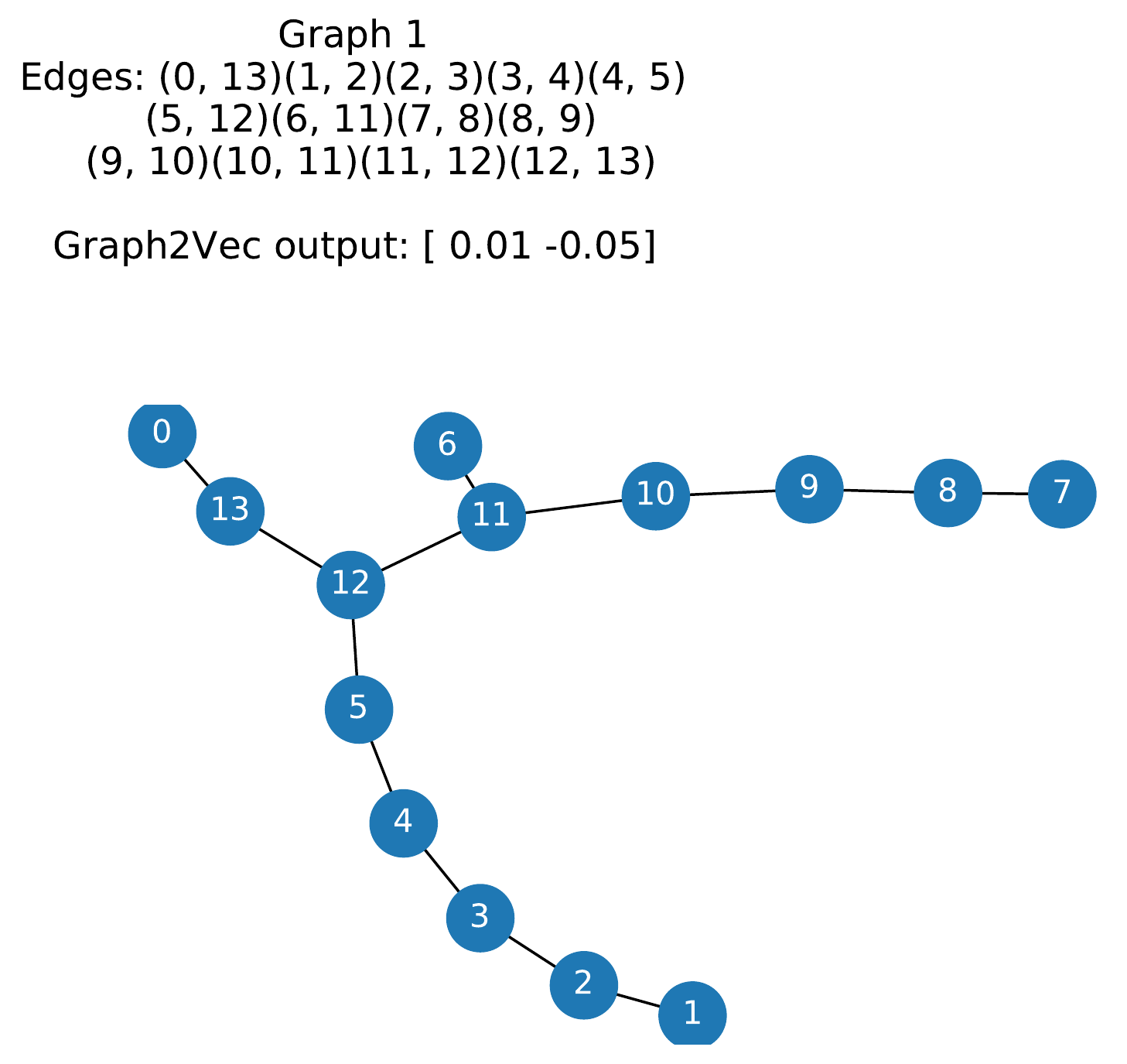}
	\caption{Converted the DepGraph shown in Figure~\ref{php2} into an embedding using Graph2Vec.}
	\label{fig:graph2vec}
\end{figure}

To show an example of converting a DepGraph into a graph embedding using Graph2Vec, we converted the DepGraph of the previous example, shown in Figure~\ref{php2}. Figure~\ref{fig:graph2vec} displays the input edges and nodes and the corresponding embedding output. It is worth noting that similar graphs have similar vector representations, and identical graphs have similar but not identical vector representations since embeddings are randomly initialized. This shortcoming is not a concern for our approach since the strong similarity is enough. Additionally, Graph2Vec requires graphs with undirected edges, while DepGraphs have directed edges. Therefore, to use Graph2Vec with our dataset, we need to convert all the directed edges into undirected edges, losing some information in the process. Nonetheless, Graph2Vec is suitable for our method since it preserves most of the relevant information.

\subsection{Automatic Analysis}
\label{use-cases}

As explained in the previous section~\ref{graph2vec}, a graph embedding is learned for each DepGraph in the kernel trace dataset. Using outlier detection algorithms, we can automate the process of finding outliers by clustering and analyzing similar DepGraphs. After identifying the outliers, we use a comparison algorithm to compare each outlier to a cluster of normal executions to find the differences, hence the cause of the given outlier.

\subsubsection{Outlier Detection}
\label{ouliter_detection_and_analysis}

The definition of an outlier as provided by \citet{hawkins1980identification} is: ``\emph{an observation which deviates so much from the other observations as to arouse suspicions that it was generated by a different mechanism.}’'

Outliers may correspond to noise in the dataset, unusual patterns or behaviours of interest, or anomalies such as latency, intrusions, and bugs. In our paper, all the outliers detected are latency-related issues. We consider a latency outlier to be an execution/request with a high execution/response time. Most of the requests in our dataset have a runtime lower than 200ms (97.8\% of the dataset). Any request that has a runtime above 200ms is considered an outlier. Furthermore, any request with unusual internal nodes is considered an outlier, even if the execution time is lower than 200ms. After all, they represent a system resource, system call, or request that has an unusual behaviour compared to the other requests in the dataset. Using the outliers we are able to observe the potential reason for the performance issue, since it displays the internal threads/processes with high runtime or it might show unnecessary threads/process that the request is using, making it slow.

\begin{figure*}
	\centering
	\begin{subfigure}[b]{1\textwidth}
		\includegraphics[width=\linewidth]{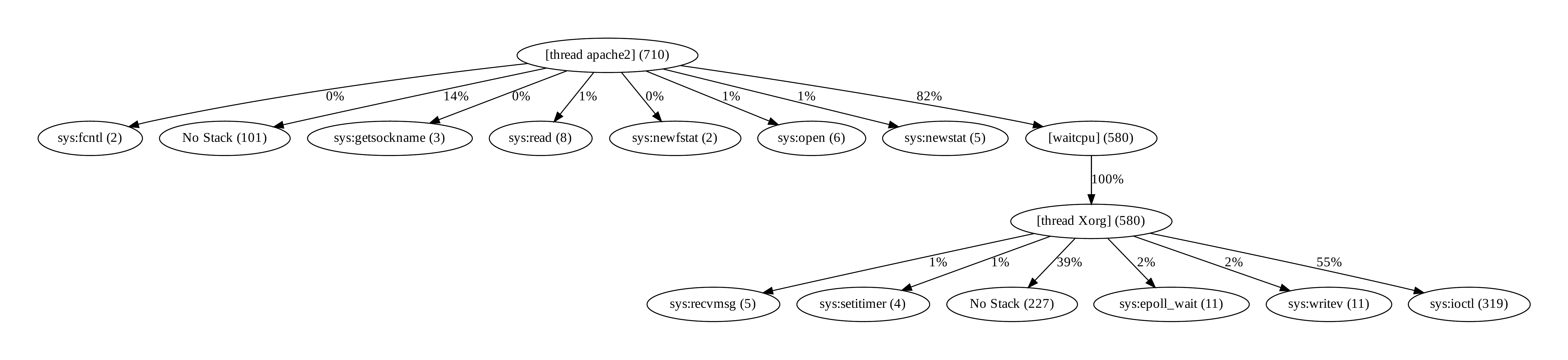}
		\caption{This DepGraph is considered to be an outlier because it contains novel elements compared to other DepGraphs in our dataset. Also it is one of the only DepGraph with that big of a request time.}
		\label{fig:outlier1}
	\end{subfigure}

	\begin{subfigure}[b]{1\textwidth}
		\includegraphics[width=\linewidth]{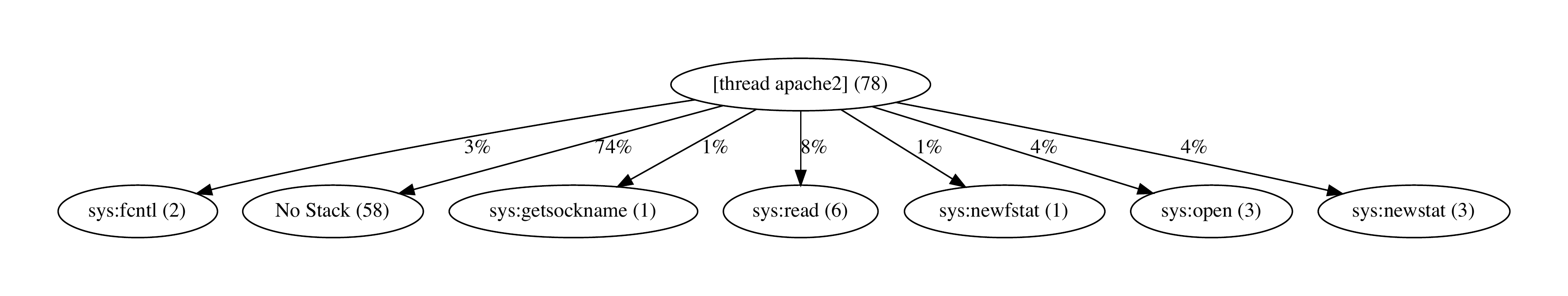}
		\caption{This DepGraph is considered to be a normal request, since it has a runtime under 200ms and it contains similar elements as the other clustered DepGraphs.}
		\label{fig:normal1}
	\end{subfigure}

	\caption[Two numerical solutions]{Displays an example of a normal request and an outlier to differentiate between the two.}
\end{figure*}

From our dataset, we show the DepGraph in Figure~\ref{fig:outlier1} as an outlier compared to the DepGraph in Figure~\ref{fig:normal1}. Figure~\ref{fig:outlier1} has a runtime of 710ms and also contains elements that are unusual compared to other DepGraphs in the dataset, such as \texttt{thread Xorg}. Meanwhile, Figure~\ref{fig:normal1} is a normal request since it only has a runtime of 78ms, and all of the elements are similar to other DepGraphs in the dataset. Later on, in Section~\ref{comparing_depgraph}, we display the use of a comparison algorithm to compare the outlier shown in Figure~\ref{fig:outlier1} with the other normal DepGraphs to find the differences between the two and find the root cause of the outlier.

We use three different models to detect outliers: $Z$-score, distance-based model, and density-based model.

The $Z$-score approach indicates how many standard deviations away a point lies from the sample's mean value. We decided to use $Z$-score because it is an effective method to detect outliers on a Gaussian distribution~\citep{santoyo_2017} and because it is easy to implement. The limitation of the $Z$-score is that it may provide inaccurate results with smaller datasets. However, the sample size would not usually be an issue for our use case since, for accurate $Z$-score calculations, the sample size should be greater than 30. $Z$-score uses the mean and standard deviation of the dataset to calculate its value. Outliers can highly influence these values. Therefore, datasets with too many outliers may also yield inaccurate results. 

For the distance-based model, we used the $k$-nearest neighbour ($k$-NN) algorithm. A data point is defined as an outlier if the distance between it and its $k$-nearest neighbour is greater than a predefined threshold. While the distance-based models are easy to understand and implement, they are sensitive to the distance metric.

Density-based models assume that outliers are points that lie in a sparsely populated region of space. We consider the Density-Based Spatial Clustering of Applications with Noise (DBSCAN~\citep{ester1996density}) and Ordering Points To Identify the Clustering Structure (OPTICS~\citep{ester2013density}) methods. Although these methods were designed for clustering, they are also able to identify outliers as a byproduct. DBSCAN and OPTICS work similarly to each other. Both algorithms require two parameters: $\epsilon$ and $m$. $\epsilon$ is the maximum distance between two data points for them to be neighbours, and $m$ is the minimum number of points required to create a cluster. The algorithms start by selecting a random point $x$, then count the points in its $\epsilon$-neighbourhood. If the number of points is greater than $m$, DBSCAN creates a new cluster with $x$ and its neighbours, whereas OPTICS creates an ordered list based on the reach-ability distance by keeping a priority queue. If the above condition is not satisfied, DBSCAN and OPTICS both consider $x$ to be an outlier. However, the neighbours of $x$ are considered and added to the cluster in DBSCAN, and OPTICS will add it to its ordered list. After this process, OPTICS and DBSCAN choose a new data point. When all data points have been considered, DBSCAN and OPTICS terminate. It is important to note that outliers declared by DBSCAN and OPTICS are only points that lie in a low-density region.

\begin{figure*}[!htb]
	\centering
	\begin{subfigure}[b]{0.5\linewidth}
		\centering
		\includegraphics[width=\linewidth]{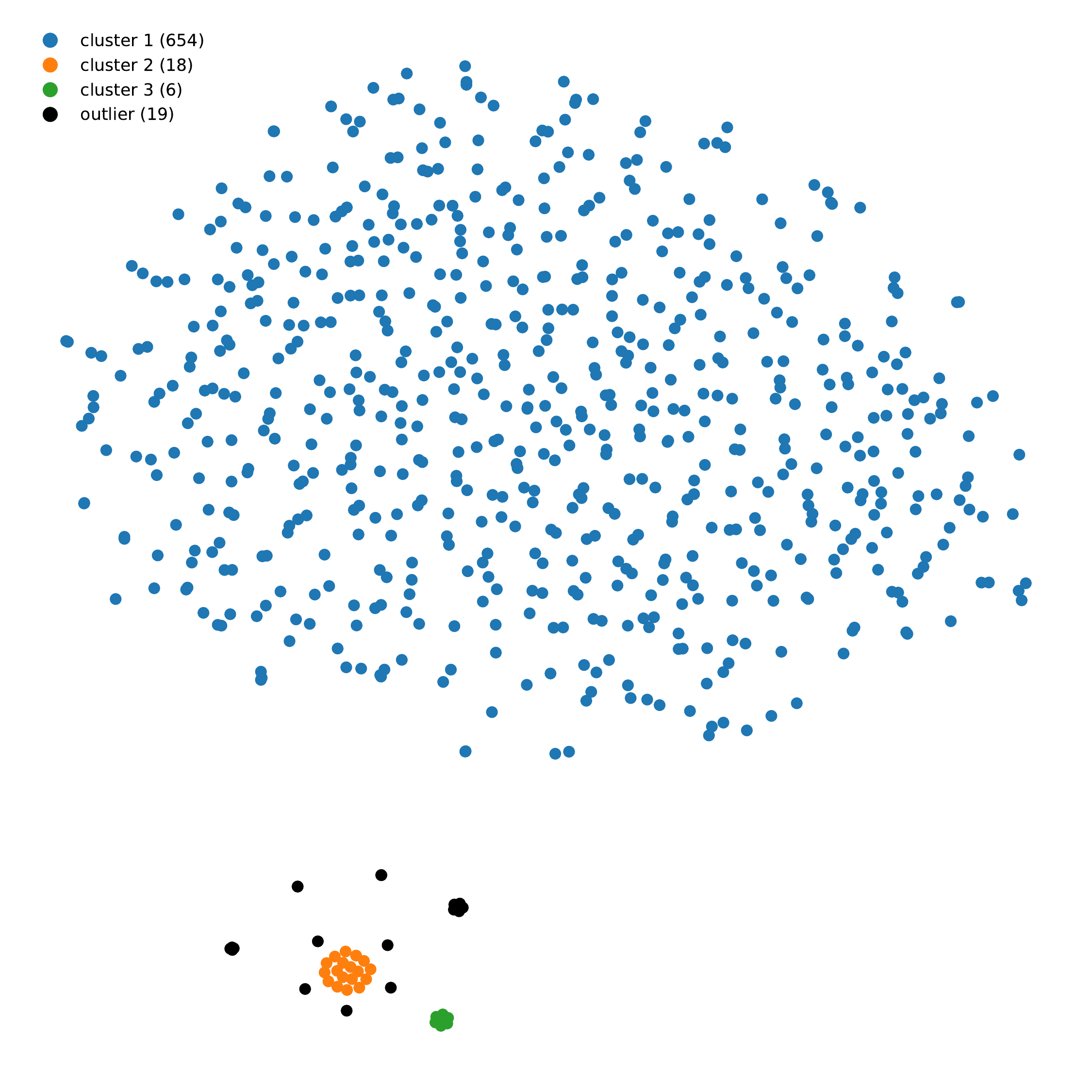}
		\caption{DBSCAN}
		\label{fig:dbscan}
	\end{subfigure}\hfill
	\begin{subfigure}[b]{0.5\linewidth}
		\centering
		\includegraphics[width=\linewidth]{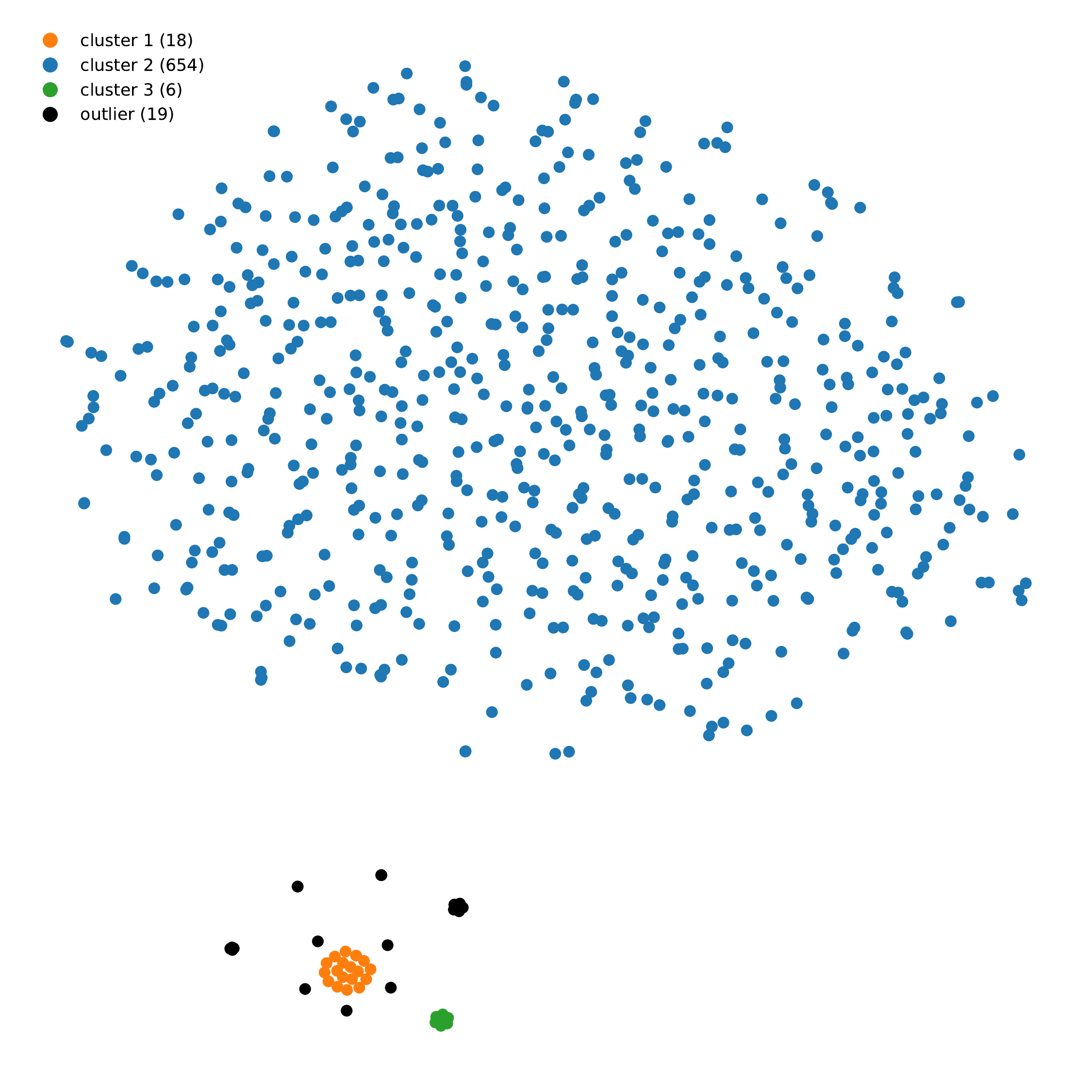}
		\caption{OPTICS}
		\label{fig:optics}
	\end{subfigure}
	\caption{Results of two different clustering methods. The visualization is obtained by projecting the output of Graph2Vec in 2D using t-SNE~\citep{JMLR:v9:vandermaaten08a}. (a) DBSCAN is a clustering method that can produce outliers as a byproduct of clustering. It is not built for outlier detection, however, it is often used for that. (b) OPTICS produces similar results to DBSCAN's since they are both density-based methods.}
	\label{fig:clustering}
\end{figure*}

Figure~\ref{fig:dbscan} and Figure~\ref{fig:optics} show a visualization of the DepGraphs clustering with OPTICS and DBSCAN. From these figures, we observe that both methods cluster points in a similar manner, as they were able to get the same clusters and outliers.

Our reasoning behind using DBSCAN and OPTICS is that we can detect outliers with great accuracy and cluster the rest of the dataset into likewise groups. The clusters created by these algorithms are later used in Section~\ref{comparing_depgraph}. In Section~\ref{comparing_depgraph} we show our comparison method to compare the outliers with the cluster of normal executions to find the difference between the two and possibly identify the root cause of an outlier.

Evaluating the outlier detection is challenging since it is an unsupervised problem. In the past, synthetically generated datasets or a few rare aspects of a real dataset were often used as proxies to evaluate unsupervised algorithms. Since external criteria such as known labels have restricted requirements, one might look for internal criteria for outlier validation. However, internal criteria are hardly ever used in outlier analysis because they are known to be flawed~\citep{aggarwal2015data}. Internal criteria's flaw is that since outliers are rare by nature, a certain validity measure might favour one outlier detection algorithm over others. Therefore most of the validity measures used in outlier analysis tend to be external measures.

Typically, unsupervised algorithms are evaluated with internal measures such as the value of a loss function on the training set. Outlier detection algorithms may also be evaluated with ground truths retrieved from synthetic or real datasets. In that case, one may see the outlier detection task as a supervised problem. Typically, soft methods return a probability of belonging to the positive class, and a simple decision threshold is used to make the classification (outlier or normal). Then, standard metrics such as accuracy, precision, and recall may be computed.

An outlier detection algorithm's primary goal is to declare all positive outliers and ensure as few false positives and false negatives as possible. In order to achieve this in a real-world scenario, one may do a grid search on the decision threshold value\footnote{A grid search exhaustively considers all parameter combinations for the set of values specified.} or a random search\footnote{A random search samples a given number of parameter combinations from the distributions specified.}.

The dataset used for this paper did not contain the ground truths. As a solution, we considered slow requests as the only outliers, although some fast requests could also contain unusual behaviours or patterns. The binary labels were generated using a response time threshold set by an expert and used to evaluate the different outlier detection methods. The threshold was 200ms since the majority of the dataset was under 200ms. Based on this threshold value 2.5\% of the dataset was declared as an outlier. 

\begin{table}[!htb]
	\centering
	\caption{Accuracy, precision, recall of each outlier detection method used.}
	\begin{tabular}{lrrrr}
		               & DBSCAN & OPTICS & $k$-NN & $Z$-SCORE     \\ \hline
		\# of outliers & 19     & 19     & 17     & 20            \\
		Accuracy (\%)  & 97.7   & 97.7   & 97.1   & \textbf{98.1} \\
		Precision (\%) & 47.4   & 47.7   & 35.3   & \textbf{55.0} \\
		Recall (\%)    & 60     & 60     & 40.0   & \textbf{73.3} \\
		F1 (\%)        & 52.9   & 52.9   & 37.5   & \textbf{62.9} \\\hline
	\end{tabular}
	\label{tab:outlieranalysis}
\end{table}

Table~\ref{tab:outlieranalysis} displays the accuracy, precision, recall, and f1 score of each outlier detection method considered (DBSCAN, OPTICS, $Z$-Score, and $k$-means). Each of these metrics requires two parameters \texttt{ground truths} and \texttt{predictions}.  While the \texttt{ground truth} remains the same for each method, the \texttt{predictions} are different depending on the outlier detection method.

The results obtained from this evaluation methodology should be considered with caution since detected outliers may correspond to anomalies unrelated to latency, in which case we acknowledge the prediction to be erroneous. The accuracy measures the ratio of correctly predicted samples, both normal and outliers, while the precision and recall only consider the outliers. The f1 score is a harmonic mean of precision and recall. The table allows us to determine which outlier detection methods were successful in detecting slow requests.

\subsubsection{Comparing DepGraphs}
\label{comparing_depgraph}

So far, using the proposed outlier detection methods, we have been able to find anomalies and group similar DepGraphs. Now using these groups and outliers, we would like to find the root cause of these anomalies. This is achieved by first merging all the DepGraphs in each cluster. By merging all the DepGraphs in a cluster, we create a new DepGraph representing the entire cluster. We can then use the comparison algorithm to discover the differences between the outliers and the merged graphs.

\begin{algorithm}
	\caption{Method used to merge the cluster}
	\label{merging_algo}
	\begin{algorithmic}[1]
		\Procedure{Merge}{$file$}
		\State $Data \gets formatData(path, function(file))$ \Comment{Collect all the executions and make sure, there are no loose nodes}

		\State $max = dictionary()$
		\State $min = dictionary()$
		\State $count = dictionary()$
		\State $size = dictionary()$

		\For{each execution in data}
		\For{each node in the given execution}
		\Comment{Add the node if and only if its distinct else add the value}
		\If {nodePath is not in counts}
		\State $count[nodePath] = 1$
		\Else
		\State $count[nodePath] = count[nodePath] + 1$
		\EndIf
		\If {nodePath is not in min}
		\State $min[nodePath] = nodePath[min]$
		\ElsIf{nodePath[min] \textless \ min[nodePath]}
		\State $min[nodePath] = nodePath[min]$
		\EndIf
		\If {nodePath is not in max}
		\State $max[nodePath] = nodePath[max]$
		\ElsIf{nodePath[max] \textgreater \ max[nodePath]}
		\State $max[nodePath] = nodePath[max]$
		\EndIf

		\If {nodePath is not in sizes}
		\State $sizes[nodePath] = nodePath[size]$
		\Else
		\State $sizes[nodePath] = sizes[nodePath] + nodePath[size]$
		\EndIf

		\EndFor
		\EndFor

		\State \Call{createGraph}{$size, count, min, max$} \Comment{Pass on the new distinct nodes to make a new DepGraph}
		\EndProcedure

	\end{algorithmic}
\end{algorithm}

Algorithm~\ref{merging_algo} displays the method used to merge the clusters. As input, it takes a file comprising of the DepGraphs to merge. After retrieving the file, the algorithm formats the data by extracting all the information of each execution and ensuring no nodes without a parent exist. Then, the algorithm analyzes each node from each execution (line 7). If the node is unseen, the dictionary \texttt{counts} would add it; otherwise, it would increase the count of that node, indicating that the node is repeated (lines 9-13). The algorithm also records the minimum and maximum values of all similar nodes. Therefore if a node’s size is less than the current size, it will be updated as the new minimum. Similarly, if a node’s size is greater than the current size, it will be updated as the new maximum. If it is a distinct node, its value becomes the maximum and the minimum (lines 14-23). Furthermore, the algorithm stores the sizes of each distinct node. If the node already exists, it adds its value with the new node’s value (lines 24-28). Using the \texttt{size}, \texttt{count}, \texttt{min}, and \texttt{max} (line 31) dictionaries, the algorithm is able to make a DepGraph that has the id and name of all distinct nodes, the cumulative size of all repeated nodes, the number of repeated nodes (count), the max and min values from all the common nodes, and the path.

\begin{algorithm*}
	\caption{Method to compare the merged cluster DepGraph and the outlier.}
	\label{compare_algo}
	\begin{algorithmic}[1]
		\Procedure{Merge}{$file1, file2$}

		\State $data1 \gets formatData(file1)$
		\State $data2 \gets formatData(file2)$
		\State $totCount = dictionary()$
		\State $means1 = dictionary()$
		\State $means2 = dictionary()$
		\State $leftCounts = []$
		\State $meanDiffSd = dictionary()$
		\State $diffMeans = dictionary()$
		\State Merge the counts, max, min and size for the \texttt{data1} and \texttt{data2} \Comment{For loop on line 7 in Merge algorithm~\ref{merging_algo}}

		\For{each node in execution1}
		\If{node[path] in totCount}
		\State $totCount[node[path]] = totCount[node[path]] + node[count]$
		\Else \
		$totCount[node[path]] = node[count]$
		\EndIf
		\EndFor

		\For{each node in execution2}
		\If{node[path] in totCount}
		\State $totCount[node[path]] = totCount[node[path]] + node[count]$
		\Else \
		$totCount[node[path]] = node[count]$
		\EndIf
		\EndFor

		\Comment{mean of each node count is count/totalcount for that node}
		\For{node in execution1}
		\State $means1[node[path]] = node[count] / totCount[node[path]]$
		\State $leftCounts.append(node[count])$
		\EndFor

		\State $leftSds = standardDeviation(leftCounts)$

		\For{node in execution2}
		\State $means2[node[path]] = node[count] / totCount[node[path]]$
		\EndFor
		\For{mean in means1}
		\If{mean is in means2}
		\State $diffMeans[means] = means[mean] - means[mean]$
		\EndIf
		\EndFor

		\For{diff in diffMeans}
		\If{leftSds != None}
		\State $meanDiffSd[diff] = diffMeans[diff] / leftSds$
		\Else
		\State $meanDiffSd[diff] = infinity$
		\EndIf
		\EndFor

		\For{x in meanDiffsd}
		\State $boldness[x] = getBold(meanDiffSd[x])$
		\EndFor

		\Call{createGraph}{$Merged min, Merged max, Merged count, Merged size, boldness$}
		\\\Comment{since the comparison graph has all nodes from both depgraphs it needs the merged data we got at line 10}
		\EndProcedure
	\end{algorithmic}
\end{algorithm*}

We then use the comparison algorithm shown in Algorithm~\ref{compare_algo} to compare the merged graph (i.e., the representative graph of the normal cluster) and the DepGraph of the outlier request. As input, the algorithm takes two files, each containing a DepGraph to compare. In our case, one DepGraph represented the clusters, and the other represented the outlier. After retrieving both files, the algorithm extracts all the nodes from both executions and merged their data (\texttt{counts}, \texttt{max}, \texttt{min}, and \texttt{size}) using a similar method shown in Algorithm~\ref{merging_algo}. Next, the algorithm gets the total number of repetitions for each node from both executions (lines 12-22). Using the \texttt{totCount}, the algorithm calculates the mean count of each node in the first execution and calculates the standard deviation of the first DepGraph (\texttt{leftSds}). Later it repeats the process and calculates the mean count of each node in the second execution (lines 23-30). With the two means, the algorithm can find the differences between the means, which is needed to find the similarities between the two DepGraphs. The algorithm iterates over each difference in \texttt{diffMeans} and calculates the difference between the standard deviation of the means. If the two nodes are present in both the DepGraphs, the algorithm uses \texttt{getBoldness} method to compare each node’s count. Depending on the standard deviation, it can determine the level of boldness. The higher the standard deviation, the higher the difference between the counts of the two nodes. We use boldness to show the differences in the value of the corresponding nodes of the given graphs. The \texttt{getBoldness} method uses five different boldness levels. The larger the value difference, the higher the boldness level of the edges. The algorithm can also show distinct nodes that may only appear in one of the two executions. If the node is only present in the first DepGraph, it will have a dashed edge, and if the node is only present in the second DepGraph, it will have a dotted edge. The resulting graph will depict the exact differences (i.e., value differences between corresponding nodes and existence of a node in only one tree) between the two given DepGraphs, useful in identifying the root cause(s) of the outlier request.

\begin{figure*}[htbb]
    \centering
	\includegraphics[width=\linewidth]{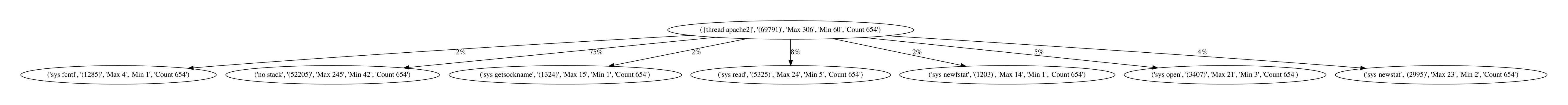}
	\caption{This DepGraph is a result of merging all the DepGraphs in cluster 2 from the OPTICS clustering method.}
	\label{fig:optics_merge_cluster1}
\end{figure*}

\begin{figure*}[htbp]
	\centering
	\includegraphics[width=1\linewidth]{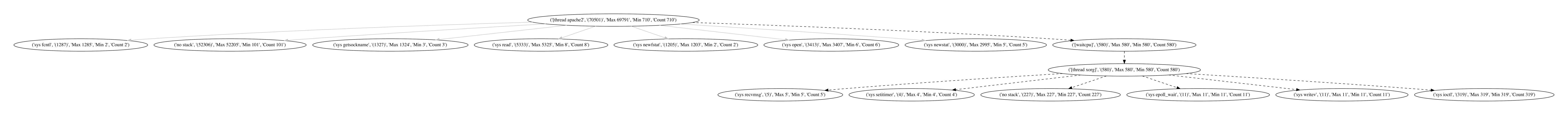}
	\caption{This is a comparison graph that compares the cluster 2 from OPTICS and the outlier shown in Figure~\ref{fig:outlier1}}
	\label{fig:comparision_graph}
\end{figure*}

To display an example of using the merge and comparison algorithms to find root causes of anomalies, we will compare the outlier shown in Figure~\ref{fig:outlier1} and the normal DepGraph in Figure~\ref{fig:normal1}. Using OPTICS, we can cluster the normal DepGraph in cluster 2 and identify the outlier. Before comparing the cluster with the outlier, we first need to merge the entire cluster into one DepGraph since the comparison algorithm can only take two DepGraphs as input. Using the Algorithm~\ref{merging_algo}, we merged cluster 2 into a single representative DepGraph with all distinct nodes. Figure~\ref{fig:optics_merge_cluster1} displays the final DepGraph after merging all 654 (normal) DepGraphs in the cluster. In this figure, each node has a name, the total cumulative size of all the same nodes, the maximum and minimum of all the same nodes (i.e., the execution/run time of that node), and the number of repeated nodes. For example, in the DepGraph, the node \texttt{thread apache2} shows a count of 654, which means that 654 DepGraphs had the same node in their DepGraphs. Out of the 654 nodes, the highest value was the runtime of 306ms, and the lowest value was the runtime of 60ms. The node also shows a cumulative size of 69,791, which allows computing the percentage of time spent by the thread in the node. With the single merged DepGraph that represents cluster 2, we can now compare it with the outlier in Figure~\ref{fig:outlier1} using the comparison Algorithm~\ref{compare_algo}. Using the Algorithm~\ref{compare_algo}, we get the comparison graph shown in Figure~\ref{fig:comparision_graph}.

The comparison graph shown in Figure~\ref{fig:comparision_graph} depicts all distinct and common nodes from both of the DepGraphs. Nodes with dotted edges are from the first DepGraph (representing cluster 2), and nodes with dashed edges are from the second DepGraph (outlier). If an edge is a simple straight line (possibly with boldness level), it represents that the node is present in both the DepGraphs. In Figure~\ref{fig:comparision_graph}, we can observe that most nodes are present in both DepGraphs. However, \texttt{waitcpu} is only present in the second DepGraph (outlier). By investigating the \texttt{waitcpu}, we can identify nodes (e.g., other active thread(s), which is the thread \texttt{xorg} here) that take up the majority of the CPU time while the main thread is waiting to get the CPU. In other words, this dash-lined node from the difference graph shows that the CPU was utilized by \texttt{xorg} thread, and consequently, it is not available for the apache2 thread to handle the web request causing it to get blocked and wait longer than usual to get the CPU. We can conclude that this is one of the reasons why this request is an outlier. There might be, however, some other reasons that can be seen from the difference graph.

\section{Evaluation}
\label{general_disscussion}

\subsection{Computational Cost} \label{cost}
\subsubsection{Setup}

The trace used for our example above was collected using LTTng 2.11 on a workstation equipped with 32 GB of RAM and a quad-core Intel\textregistered~Core\texttrademark~i7-6700K CPU @ 4.00 GHz. The operating system was Linux Ubuntu 16.04.6 LTS with the 64-bit kernel 4.15.0-62. The local hard disk reference 7200 RPM WDC WD1003FZEX-0 stored the trace data and trace analysis results.

\subsubsection{Data}

\begin{figure}[htb]
	\centering
	\includegraphics[width=\linewidth]{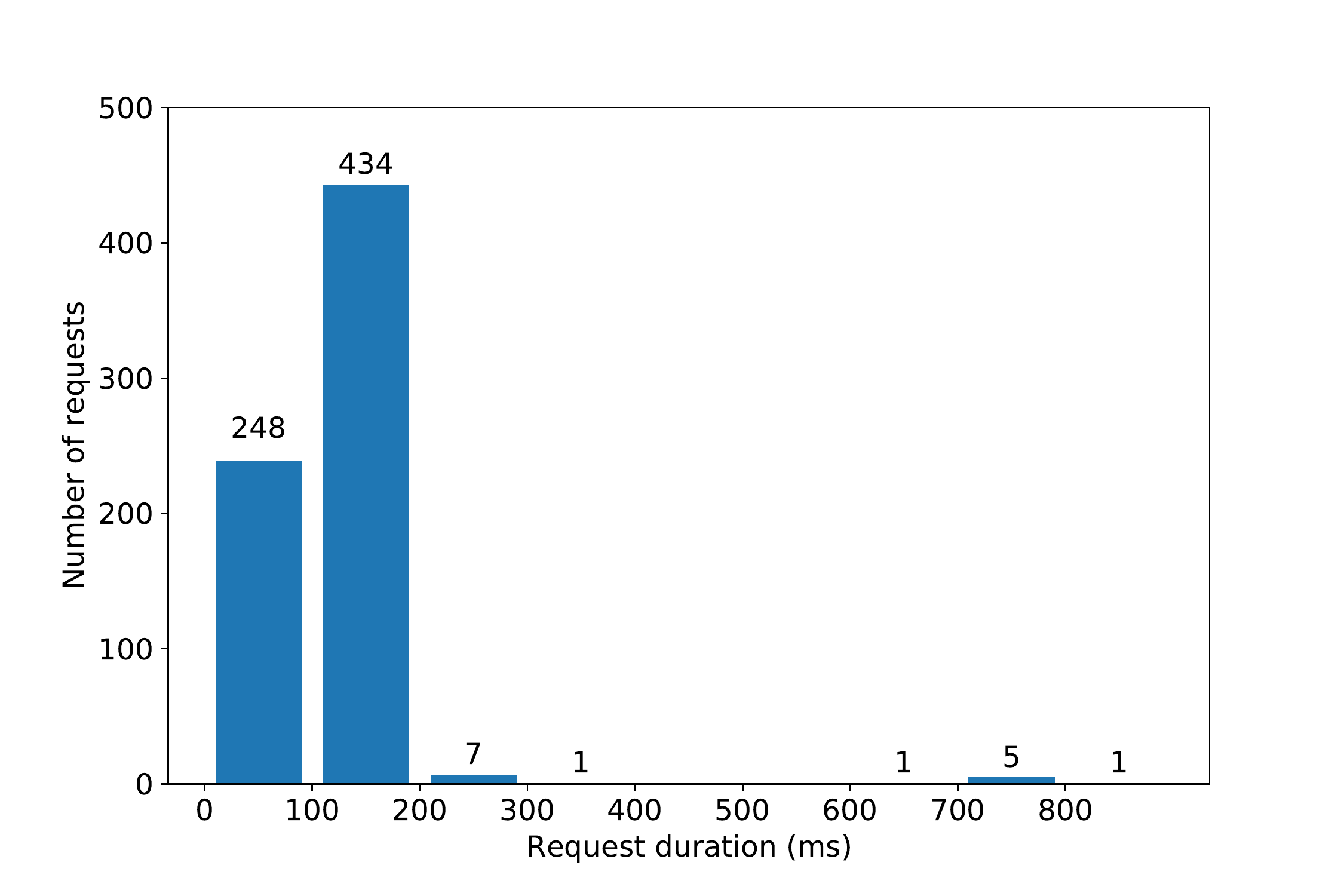}
	\caption{Histogram of the request duration in ms.}
	\label{fig:req_graph}
\end{figure}

The trace data was collected using the LTTng tracer and comprised 697 requests. The trace contains the execution of an Apache web server handling a PHP web application. The internal execution of each request is reflected in a distinct DepGraph. Therefore, there are 697 DepGraphs in the dataset. Our observation shows that 97.8\% of all requests have a runtime between 0 and 200ms (considered as normal executions), as shown in Figure~\ref{fig:req_graph}. Therefore, for this dataset, we consider any request and corresponding DepGraph that has a runtime below 200ms to be a normal request and the rest to be slow requests. Normal requests can still contain anomalies that are unrelated to latency issues. More details of the dataset and the constructed DepGraphs can be found in \citet{ezzati2020depgraph}.

\subsubsection{Tracing and DepGraph Construction Cost}

In the worst-case scenario of tracing the Linux kernel where all kernel tracepoints are enabled, which is not the case of our method, the overhead imposed by tracing is 42\%. In our method, when collecting the trace to construct DepGraphs, only system calls and events needed to extract the DepGraphs are enabled. Our analyses show that the overhead imposed by the tracing never exceeds 10.1\%. After obtaining the trace, the DepGraphs were created. When creating the DepGraphs, one needs to analyze the trace, which is done offline after collecting it. Therefore it does not have any overhead on the application and is not considered in our time analysis.

\subsubsection{Automated-Analysis}

\def\arraystretch{1.25}
\setlength\tabcolsep{8pt}
\begin{table*}[!htb]
	\centering
	\caption{Time taken by each task.}
	\begin{tabular}{llrr}
		                                              &                                  & Time (s) \\ \hline
		\multirow{2}{*}{Data Processing}              & Load DepGraphs                   & 0.076    \\
		                                              & DepGraph Embeddings w/ Graph2Vec & 11.608   \\ \hline
		\multirow{2}{*}{Cluster}                      & DBSCAN                           & 0.020    \\
		                                              & OPTICS                           & 0.325    \\ \hline
		\multirow{4}{*}{Outlier Detection}            & DBSCAN                           & 0.017    \\
		                                              & OPTICS                           & 0.275    \\
		                                              & $k$-NN                           & 0.075    \\
		                                              & $Z$-Score                        & 0.025    \\ \hline

		\multirow{3}{*}{Merging clusters}             & Load the cluster                 & 0.027    \\
		                                              & Merge DepGraphs in cluster       & 0.033    \\
		                                              & Construct DepGraphs              & 32.391   \\ \hline

		\multirow{3}{*}{Compare outlier with cluster} & Load the two DepGraphs           & 0.001    \\
		                                              & Merge both execution for details & 0.001    \\
		                                              & Compare the two DepGraphs        & 0.002    \\
		                                              & Construct comparison  graph      & 0.672    \\
		\hline
	\end{tabular}
	\label{tab:timetable}
\end{table*}

Table~\ref{tab:timetable} displays the time taken for various tasks in Section~\ref{use-cases}. The table shows the time it took to load the dataset, make the graph embeddings, and run each outlier detection and clustering method. From that, we observed that DBSCAN took the least amount of time to cluster and detect outliers. The table also shows how long it took to merge a cluster into one DepGraph and compare it with an outlier. For this table, we measured the time taken to merge cluster 2 and the time taken to compare it with the outlier shown in Figure~\ref{fig:outlier1}. Here we notice that the majority of the time was spent constructing the DepGraphs and embedding, whereas everything else took less than a second to do.

The overhead imposed by our method is modest compared to the cost of tracing and certainly negligible compared to the time required by a human operator to investigate each request manually.

\subsection{Limitation}
\label{limit}

The proposed method described in this paper is heavily reliant on Graph2Vec used to get our fixed-size representation of a trace. However, it has a few limitations. When training the Skip-gram model to produce the embeddings, it is primarily reliant on the embedding size and the hyper-parameters \texttt{wl\_iterations} and \texttt{epochs}. Therefore when we increase the epochs and iterations, the time it takes to train the model also increases. If a significant portion of the graphs used to train the model is large, that can also increase the time it takes to train the model. Also, to train the model to embed, a minimum number of graphs are needed. If there is less than the minimum threshold, we run into a small sample problem, resulting in overfitting.

Another limitation our proposed method has is that it can only detect off-CPU issues and cannot detect any issues that happen at the on-CPU level. Since our proposed method focuses on finding latency issues within dependency graphs, this is not an issue. However, one cannot use our method to detect issues within their code (on-CPU analysis). To do so, we would need to get data from on-CPU profilers. Since there are no interactions between kernel space and user space, one cannot get information about the user space functions from the kernel level.

\section{Conclusion and Future Work}
\label{conclusion}

Our proposed method is able to automatically detect outliers and their root causes from within a system-level trace data. Using our method, we remove the need to manually examine all the trace data and the abstract data (such the dependency graphs) extracted from them. Furthermore, with our approach to compare the declared outliers with the normal executions, we can identify unusual activities contributing to the latency. This method makes the process of analyzing the software execution trace more efficient for developers, as automating the process reduces the time taken for debugging and gets the task done to find potential bottlenecks in the program. Additionally, having an automated process removes the potential for human error in the analysis. For future work, we would like to extend our approach by reducing the graph embedding size while still retaining the level of accuracy (or potentially increasing it). By reducing the dimension of the vectors, we can train the models faster while increasing the number of dependency graphs.

\bibliographystyle{IEEEtranN}
\bibliography{bibliography}

\end{document}